\documentstyle[twocolumn,epsfig,aps]{revtex}
\draft

\begin{document}
\twocolumn[\hsize\textwidth\columnwidth\hsize\csname
@twocolumnfalse\endcsname

\title{Elastic properties of 2D colloidal crystals from video microscopy}

\author{K. Zahn, A. Wille and G. Maret}

\address{$^{1}$Universit\"at Konstanz, Fachbereich f\"ur Physik,
POB 5560, D-78457 Konstanz, Germany}

\author{S. Sengupta$^{2}$ and P. Nielaba$^{1}$}

\address{$^{2}$Satyendra Nath Bose National Centre for Basic Sciences,
Block JD, Sector III, Salt Lake, Kolkata 700 098, India}

\date{\today}

\maketitle

\begin{abstract}
Elastic constants of two-dimensional (2D) colloidal crystals are
determined by measuring strain fluctuations induced by Brownian
motion of particles.
Paramagnetic colloids confined to an air-water interface of a
pending drop are crystallized under the action of a magnetic
field, which is applied perpendicular to the 2D layer. Using
video-microscopy and digital image-processing we measure fluctuations
of the microscopic strain obtained from random displacements of 
the colloidal particles from their mean (reference) positions. 
From these we calculate system-size
dependent elastic constants, which are extrapolated using finite-size 
scaling to obtain their values in the thermodynamic limit.
The data are found to agree rather well with zero-temperature calculations.
\end{abstract}

]
%elastic constants 62.20.Dc
%Colloids 82.70.Dd

%\begin{multicols}{2}
%\narrowtext
During the last two decades interest in colloidal systems has
grown substantially, on one hand because of their widespread
technological applications and on the other due to the
availability of precisely calibrated particles for use as model
systems for studying phenomena in classical
condensed matter physics\cite{pus91}. The crystallization of
colloids, both in two and three dimensions has been a continuous
matter of interest. The research mostly focused on the analysis of
structure and dynamics of colloidal systems on different length
and time scales through static or dynamic light scattering
techniques. Measurements of elastic constants of colloidal
crystals, however, have been limited to the determination of the shear
modulus $\mu$. This was based on the observation of shear induced
resonance of a crystal using light scattering techniques (see
\cite{sch98} for a recent work). The value of $\mu$ is found to
depend strongly upon the crystalline morphology and changes
significantly between randomly oriented crystallites and
shear-ordered samples \cite{pal94}. In addition, using this method
only a very reduced number of modes can be investigated. Very recently,
the elastic moduli of colloidal solids have also been estimated\cite{wil01} by 
observing relaxation behaviour after deformations using laser tweezers. 
\begin{figure}
\centerline{\epsfig{figure=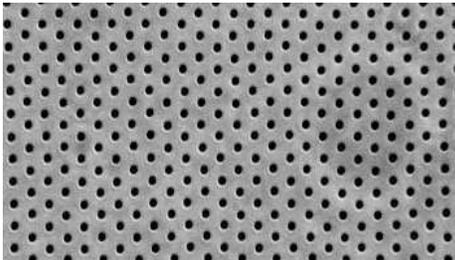, width=6.0cm}}
\caption{A snapshot of the triangular lattice of paramagnetic colloidal 
particles. A few thousand snapshots such as this, taken at regular time 
intervals of about one second were used to calculate elastic constants.}
\label{fig0}
\end{figure}

\noindent
In this Letter, we report an experimental determination of  the equilibrium
elastic properties of two~-dimensional (2D) colloidal crystals from
``snapshots'' of particle positions obtained using video microscopy.
The present method is completely non-invasive, accurate and free from any 
adjustable parameters.
\begin{figure}  
\centerline{\epsfig{figure=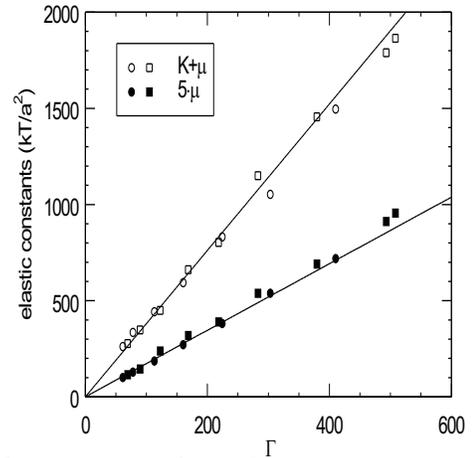,width=6.0cm,height=6.0cm}}
\caption{The comparison of the measured elastic constants (in
units of $kT/a^2$, $a$ is the lattice parameter) to the zero-temperature calculation (solid
lines) reveals an very good agreement. Note that the shear modulus
$\mu$ is multiplied by 5 for reasons of clarity.}
\label{fig1}
\end{figure}

The mechanical properties of a macroscopic solid, according to classical 
elasticity,  are well described by a small set of
elastic constants. These can be measured by the strain-response of
the solid under the application of an appropriate (macroscopic)
stress. On a mesoscopic scale, which is still sufficiently
coarse-grained to apply elasticity theory but small enough such
that the Brownian motion of the particles is observable,
thermally induced strain fluctuations can be used to determine
elastic constants. We have carried out a detailed study of these strain 
fluctuations
in a two~-dimensional (2D) colloidal crystal by
recording the microscopic positions of the particles within a square 
cell (of size $L$) containing a defect~-free single crystal 
(see Fig. \ref{fig0}). Recordings were
made at regular time intervals --- large compared to typical
correlation times of the colloid. Uncorrelated snapshots
obtained are analyzed to calculate the average particle
positions --- the ``reference'' lattice. The gradients
(obtained by finite differences) of the displacement
vectors then yield the microscopic strains. These microscopic
strains are used to obtain strain fluctuations over a hierarchy
of length scales corresponding to smaller {\em sub}-cells of size 
$L_b < L$ contained within our cell. The width of the probability 
distributions of the strains are related to the elastic constants
$C_{ijkl}(L_b)$ obtained as a function of $L_b$ which may subsequently 
be extrapolated, using a systematic finite size scaling analysis\cite{sen00a}.
The macroscopic ($L \to \infty$) values of these quantities, thus obtained, are compared to theoretical predictions {\em without any fitting 
parameters}. 
The central result, the bulk ($K$) and shear ($\mu$) elastic 
moduli are shown as a function of the interaction strength $\Gamma$ for 
our colloidal system (see below) in Fig. \ref{fig1}. 

Our experimental setup\cite{zah97,zah99} is composed of super~- paramagnetic spherical
colloids \cite{dyn} of diameter \mbox{$d=4.5\,\mu\mbox{m}$} and
mass density $1.7\;\mbox{kg/dm}^3$. They are confined by gravity
to a water/air interface, which is formed by a cylindrical drop
suspended by surface tension in a \mbox{top-sealed} ring. The
flatness of the water-air interface (\mbox{$\O=8\;\mbox{mm}$}) is
controlled within \mbox{$\pm 1\;\mu\mbox{m}$}\cite{zah97}. For
weak magnetic fields $B$ applied perpendicular to the interface
the induced magnetic moment $M$ depends linearly on $B$,
i.e.~\mbox{$M=\chi B$} with an effective magnetic susceptibility
$\chi$\cite{zah97}. The repulsive magnetic dipole-dipole potential,
between particles $i$ and $j$ separated by a distance ${\bf r}_{ij}$, 
$V(r_{ij}) = \Gamma r_{ij}^{-3}$
dominates the interaction and is absolutely calibrated by the
interaction strength \mbox{$\Gamma=(\mu_0/4\pi) (\chi B)^2 (\pi
n)^{3/2}/kT$} where $n$ denotes the 2D volume-fraction of the
particles, $k$ is the Boltzmann constant, $T$ is the ambient temperature 
and distances, ${\bf r}_{ij}$, are in units of the mean interparticle spacing.

The experiments were carried out as follows: At high $\Gamma$, in
the crystalline phase, the system was equilibrated by the application
of small AC magnetic fields in the plane of the particles. Eventually a
defect-free crystal containing several thousand particles is obtained.
The entire sample consists of approximately $10^5$ particles.
The coordinates of typically 1000 particles in the (square) field of view were
recorded in time. About 500 to 1000 independent coordinate sets, each about
a second apart, are 
necessary to obtain sufficient statistics. 
The time interval corresponds to diffusion over
about 1 micrometer which is about one pixel ---- the limit of resolution of 
our digital camera. The results do not vary significantly over the range 
of the number of data sets used.

After the
determination of the mean position of each particle ${\bf R}^0$ (taken over
the entire set)\cite{rem1} the instantaneous displacement
$\mathbf{u}(\mathbf{R}^0) = {\bf R} - {\bf R}^0$ from the mean position
was calculated for each frame and for all particles. The set of coordinates
${\bf R}^0$ constitutes, therefore, our reference lattice and the fluctuating
displacement variable ${\bf u}({\bf R}^0)$ is known at every reference lattice
point. Once the displacements are known, 
the elements of the elastic strain tensor $\epsilon_{ij}$ and the local 
rotation $\theta$ may be defined as 
gradients of the displacement ${\bf u}$ over the spatial coordinates ${\bf r}$.
\begin{eqnarray}
\label{straintens}%
\varepsilon_{ij}=\frac{1}{2}\left( \frac{\partial u_i}{\partial
r_j} +\frac{\partial u_j}{\partial r_i} \right) \;\;\; ; \;\;\;
\theta=\frac{1}{2}\left( \frac{\partial u_y}{\partial r_x}
-\frac{\partial u_x}{\partial r_y} \right),
\end{eqnarray}

\noindent
When the derivatives are replaced by appropriate finite differences, these 
quantities may be evaluated at every reference lattice point ${\bf R}^0$.
In order to evaluate elastic constants, the microscopic strains and 
rotations need to be coarse-grained or averaged over a sequence of sub-systems
(obtained in our case by dividing our square cell into integral numbers 
of smaller square sub-cells) of size $L_b = L/b$, ($b = 4,5,..,18$)
to obtain size dependent strains 
and rotations $\epsilon^b_{ij}(L_b)$ (and $\theta_b(L_b)$) 
\begin{equation}
\label{straintensmean}%
\varepsilon^b_{ij}(L^b),\theta_b(L_b) =\frac{1}{L_b^2}\int_{L_b}\varepsilon_{ij}(\mathbf{r}),\theta({\bf r})\,d\mathbf{r}.
\end{equation}

\noindent
The fluctuations
of these quantities for a particular set of sub-systems (all of size $L_b$)
over the different snapshots yield elastic constants for the size $L_b$ 
from standard thermodynamic relations.
Although this 
procedure has been used\cite{sen00a} for obtaining  
elastic moduli of the hard and soft disk systems from computer 
generated configurations, we find that some essential modifications
are required before the methods of Ref. \cite{sen00a} can be applied 
to our system. In a typical computer 
simulation\cite{sen00a,par82,woj88,sen00b,wal58,lan86} the system is 
placed in a hard constraint determined by the ensemble used. For 
example in a constant (zero) strain ensemble\cite{sen00a} the displacement 
field ${\bf u}({\bf r})$ strictly vanishes at the boundary so that the strain,
integrated over the entire system is strictly zero.  
Our experimental cell, on the other hand,  is a small region 
embedded within a much larger 
crystal. The only constraints which are appropriate for this case is that 
the displacement 
${\bf u }$ are continuous throughout the system and ${\bf u}({\bf r}) \to 0$ 
as ${\bf r} \to \infty$. In the limit of linear elasticity for our system where 
the elastic correlation length is assumed to be much smaller\cite{sen00a} than 
$L_b$, our problem reduces to the 
determination of the total energy of a single, independent,  strain 
fluctuation in an elastic medium subject to these constraints. 

Consider, therefore,  a region of size $L_b$, with a fixed constant strain 
(or rotation) $\epsilon^b_{ij}(L_b)$ ($\theta_b(L_b)$) embedded in an 
(infinite) elastic continuum. This general problem has been 
studied in detail in several standard texts on classical theory of 
elasticity\cite{lan86,hah99}.
Recall that a 2D hexagonal lattice has isotropic elastic
behavior\cite{lan86} and, therefore, can be completely described by two
independent elastic constants, which we choose to be the bulk
modulus $K$ and the shear modulus $\mu$. While the former is related to  
the fluctuations of the volume of the sub-systems
the latter may be evaluated from the fluctuations of the angle of rotation
of the sub-systems. Let us focus on a small sub-cell within our cell.
Consider for the moment that we have a disk of radius $R_b$ 
for simplicity, the final results will be cast in a 
form independent of the shape of the sub-system. Within this disk, the 
strains are given by their values which are the averages over the area of 
the disk. This disk is embedded in an infinite elastic medium with elastic
moduli $K$ and $\mu$. 

We consider first a homogeneous expansion (or compression) of the disk
by \mbox{$R_b\rightarrow R_b+\Delta r$}. The corresponding
radial-displacement $u_r$ is given as,
\begin{eqnarray}
u_r & = & \Delta r \cdot r/R_b ,\,\,\,\, r\,\, < R_b \nonumber \\
    & = & \Delta r \cdot R_b/r ,\,\,\,\, r\,\, > R_b 
\label{ucom}
\end{eqnarray}
The angular part $u_\varphi = 0$ by 
symmetry. The displacements \mbox{$(u_r,u_\varphi)$} are related
to the strain tensor by the following equations \cite{hah99}:
\begin{eqnarray}
\varepsilon_{rr} &=& \frac{\partial u_r}{\partial r}\;\;\; ;
\;\;\;
\varepsilon_{\varphi \varphi} = \frac{u_r}{r}+\frac{1}{r}
\frac{\partial u_\varphi}{\partial \varphi} \nonumber\\%
\label{displstrain}%
2\,\,\varepsilon_{r \varphi} &=&\frac{1}{r}\frac{\partial
u_r}{\partial \varphi}+ \frac{\partial u_\varphi}{\partial
r}-\frac{u_\varphi}{r}
\end{eqnarray}
Making use of the (quadratic) free energy density $f$ of the elastic
continuum,
\begin{eqnarray}
f & = & \frac{1}{2} [K(\varepsilon_{rr}+\varepsilon_{\varphi 
\varphi})^2+\mu\lbrace(\varepsilon_{rr}-\varepsilon_{\varphi \varphi})^2 
 + 4 \varepsilon_{r \varphi}^2 \rbrace] 
\label{engdens} 
\end{eqnarray}

\noindent
and integrating $f$ over the {\em entire} space, both within and outside
the disk (to account for the deformation of the surrounding medium) and using
the strains calculated from 
Eqs.\,\,\,\ref{ucom} and \ref{displstrain}, we get 
the energy \mbox{$E=2\pi(K+\mu)\Delta r^2$} necessary to
expand the disk by $\Delta r$. We may, now,  eliminate the shape dependent 
prefactors by using the volume
\mbox{$V=\pi R_b^2$} and the volume-change \mbox{$\Delta V=2\pi
R_b \Delta r$} of the disk, to obtain finally, the energy
\mbox{$E=(K+\mu)\Delta V^2/2V$}. Using the equipartition theorem
we have therefore,  
\begin{equation}
\label{vfluc}%
\langle (\Delta V_b)^2\rangle / V_b=kT/[K(L_b) + \mu(L_b)],
\end{equation}
relating the fluctuation of the volume $V_b = L_b^2$ of the sub-cell
to the sum of the bulk and shear moduli. The above relation together with 
$\Delta V_b/V_b = \varepsilon^b_{xx}(L^b) + \varepsilon^b_{yy} (L^b)$
may now be used to obtain $K(L_b) + \mu(L_b)$ for our sub-cells. 

A similar treatment leads to a relation between $\mu$ and the
local rotation of the system $\theta$ (Eq.\ref{straintens}). The
rotation of a disk of radius $R_b$ by an angle $\theta$ leads to
an angular-displacement \mbox{$u_\varphi(r)=\theta \cdot R_b^2/r$}
for $r > R_b$ ($u_\varphi(r)=0$ for $r < R_b$). 
Applying Eq.\,\,\ref{displstrain} and integrating the energy density
(Eq.\,\,\ref{engdens}) leads to the total energy for the rotation
\mbox{$E=2\pi \mu \theta^2 R_b^2$}. Equipartition then yields,
\begin{equation}
\label{mu}%
\mu = \frac{kT}{V_b} \frac{1}{\langle(2 \theta_b )^2\rangle}.
\end{equation}
This equation has precisely the same structure as Eq.\,\,\ref{vfluc}
and, therefore, similar finite size scaling schemes can
be applied. Thus Eq.\,\,\ref{mu} together with Eq.\,\,\ref{vfluc} 
enable the determination of the elastic constants of the system. 
\begin{figure}  
\centerline{\epsfig{figure=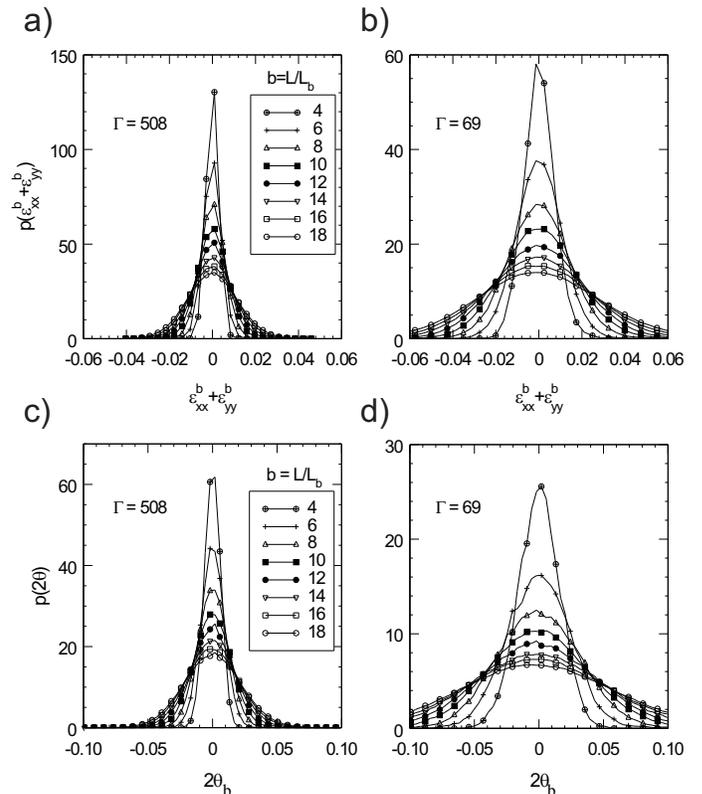,width=9.0cm}}
\caption{(a)and (b)Probability distribution of the box-size
averaged strain fluctuations of
$\Delta V_b/V_b = \varepsilon^b_{xx}+\varepsilon^b_{yy}$ as a function
 of the (linear) scaling box-size \mbox{$L_b = L/b$} for $\Gamma = 508$ (a) 
and $69$ (b). (c) and (d) Same as in (a) and (b) for the angle 
$\theta_b = \varepsilon^b_{xx}-\varepsilon^b_{yy}$ for $\Gamma = 508$ (c)
and $69$ (d). Note the difference in the y-scale between (a) and (b) as 
well as between (c) and (d)} 
\label{fig2}
\end{figure}

In Fig.\ref{fig2} the probability distribution of the
incremental volume 
\mbox{$ \Delta V_b/V_b = \varepsilon^b_{xx}+\varepsilon^b_{yy}$}
and $\theta_b$ are shown as a function of the scaled box size $b=L/L_b$ 
for two values of the interaction strength $\Gamma$. 
The width of the (Gaussian) 
distributions decreases both with increasing box-size $L_b$ and
increasing value of $\Gamma$. The 
mean square fluctuations are obtained by fitting the data to a 
normal distribution and extracting the standard deviation.

\begin{figure}
\centerline{\epsfig{figure=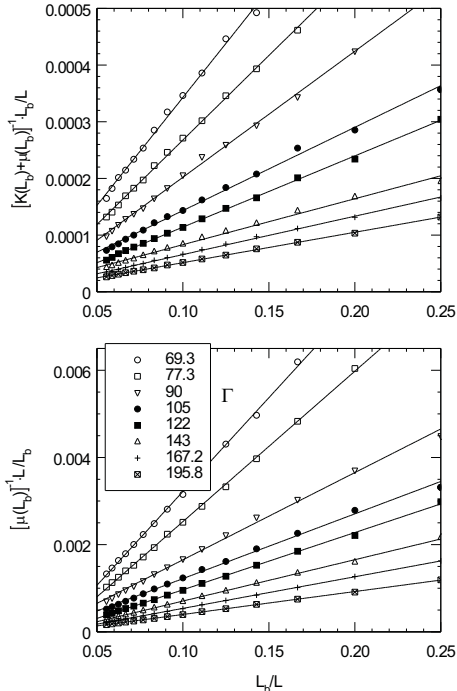,width=6.0cm}}
\caption{Finite size scaling behavior of 
$ [K(L_b) + \mu(L_b)]^{-1} \cdot L_b/L$ (top) and 
$\mu(L_b)^{-1}$ as a function of $L_b/L$
dependent on the interaction strength $\Gamma$. A straight line
fit to the curves gives the infinite system values of the
(inverse) moduli.} \label{fig3}
\end{figure}

Once the $L_b$-dependent elastic moduli are obtained, usual finite 
size scaling\cite{sen00a} can be used to extrapolate the results to the 
thermodynamic
limit. This is shown in Fig. \ref{fig3} where we have plotted $L_b/L \times 
1/[K(L_b)+\mu(L_b)]$
as a function of $L_b/L$. The slope of the curve, obtained from a 
straight line fit to the data, gives the value for 
$1/[K(L = \infty) + \mu(L = \infty)]$. A similar finite size scaling is 
used to obtain $\mu(L = \infty)$ separately.
These results are shown in
Fig.\ref{fig1} both for $K+\mu$ -- as obtained from
Eq.\ref{vfluc} -- and for $\mu$ from Eq.\ref{mu} as a function
of the interaction strength $\Gamma$. 
Note the accuracy of the determination of the elastic moduli for our
system which is facilitated by the fact 
that the interaction potential $\Gamma$ of our
system is precisely calibrated. The straight lines through our data 
in Fig.\ref{fig1} are the results from a $T=0$ calculation of the elastic
constants of a 2D triangular solid composed of particles interacting 
with an inverse cubic 
potential\cite{wildis}. For all inverse power potentials the  
$T=0$ limit is exact to lowest order\cite{invpow}. 
Considering the deformation of a perfect static triangular solid of particles
interacting with a $r^{-3}$ potential one obtains the
relation $\mu = K/10$ and the numerical result
\mbox{$K=3.461\cdot\Gamma$}, where the numerical coefficient is evaluated by 
performing a rapidly convergent lattice sum. The 
agreement is excellent (considering the fact that no fitting
parameter is available) over a wide range of interaction
strengths $\Gamma$ down to values of 70 -- the melting transition
occurs at \mbox{$\Gamma=60$} \cite{zah99}.

Finally, a few words on the possible uncertainties involved
in our determination of the elastic moduli seem to be in order. Firstly,
we have neglected all fluctuations of the magnetic moment, both in amplitude 
and 
angle. Since a super-paramagnetic colloid particle is of macroscopic dimensions 
compared to typical magnetic length scales this assumption seems to be 
justified. Secondly, we have assumed that the particles fluctuate on a 
flat, two~-dimensional air-water interface. An estimate\cite{zah97} of the 
out-of-plane fluctuations is given by the ratio of the gravitational length 
$l_g = k T/m g$ (where $m$ is the mass of the particles and $g$ is the 
acceleration due to gravity) and the interparticle spacing; this is 
typically $1$ in $10^3$. Lastly, a possible limitation of this scheme, at 
least in its present form, 
is the requirement that the displacement field {\bf u}({\bf r}) be
analytic in order to obtain strains by taking derivatives. Therefore one needs
to restrict analysis to dislocation free
regions of the sample --- which is possible only if the system is sufficiently
far away from a melting transition\cite{zah99}. Suggestions\cite{sen00a,sen00b}
to circumvent this problem are, however, computationally difficult to 
implement. 
The study of elastic properties of paramagnetic colloids in the presence 
of obstacles and inclusions, as well as dynamical elastic response is an 
interesting direction for further research. This is particularly suited 
for our technique since it provides a {\em local} (and therefore precise) 
probe\cite{wil01} for elastic properties.\\
{\bf Acknowledgements:}The authors thank M. Rao for a critical reading of
the manuscript. One of us (S.S.) thanks the Alexander von Humboldt Foundation.
Support by the Deutsche Forschungsgemeinschaft in the frame of the 
Sonderforschungsbereich 513 is kindly acknowledged.

%\end{multicols}{2}

\begin{references}
\bibitem{rem1} This is not in conflict with the divergence of the
mean square displacement in 2D crystals, which is only valid in
the thermodynamic limit. For finite systems, however, the
equilibrium position of the particles are well defined also in 2D
systems.%
\end{references}
\end{document}